\documentstyle[12pt]{article}
\begin{document}

\begin{center}

{\Large \bf The second order nonlinear conductance of a two-dimensional
mesoscopic conductor}

\bigskip

Wei-Dong Sheng$^{1,*}$, Jian Wang$^1$, and Hong Guo$^2$

\bigskip

$^1$ {\it Department of Physics, \\
The University of Hong Kong,\\
Pokfulam Road, Hong Kong.
}

\bigskip

$^2$ {\it
Centre for the Physics of Materials,\\
Department of Physics, McGill University,\\
Montreal, Quebec, Canada H3A 2T8.
}

\end{center}

\vfill

\baselineskip 15pt               

We have investigated the weakly non-linear quantum transport properties of 
a two-dimensional quantum conductor. We have developed a numerical scheme
which is very general for this purpose. The nonlinear conductance is
computed by explicitly evaluating the various partial density of states, 
the sensitivity and the characteristic potential. Interesting spatial
structure of these quantities are revealed.  We present detailed results
concerning the crossover behavior of the second order nonlinear conductance
when the conductor changes from geometrically symmetrical to asymmetrical.
Other issues of interests such as the gauge invariance are also discussed.
\vfill

\baselineskip 16pt

{PACS number: 73.20.Dx, 73.49.Ei, 73.40.Gk, 73.50.Fq }

\newpage
\section{Introduction}

Nonlinear phenomena in electric conduction play the most
important role in many electronic applications ranging from single 
units such as a diode or a transistor to entire circuits. For extremely
small systems with mesoscopic or atomic length scales, such as those 
which can now be routinely fabricated using nano-technology, quantum 
transport dominates conduction. While we now have a very good understanding 
of linear quantum transport phenomena in nano-systems where quantum 
coherence plays a vital role, the nonlinear quantum transport properties 
of mesoscopic conductors have received relatively less attention.  
In this regard, several important research results have been reported in 
recent years\cite{but1,altshu,wingreen,vegvar,tab}. 
Experimentally, Taboryski {\it et. al.}\cite{tab} have reported 
observations of nonlinear and asymmetric conductance oscillations in
quantum point contacts at small bias voltages. They found that the 
non-Ohmic and asymmetric behavior causes a rectified DC signal as the 
response to an applied AC current. On the theoretical side several
directions have been explored. Wingreen {\it et. al.}\cite{wingreen} have
presented a general formulation to deal with the situation of a non-linear
and time-dependent current going through a small interacting region 
where electron energies can be changed by time-dependent voltages. 
De Vegvar\cite{vegvar} has studied the low frequency second-harmonic transport
response of multiprobe mesoscopic conductors using a perturbation theory 
in the framework of Kubo formula and found that the low frequency
second-harmonic current is a non-Fermi-surface quantity. At the 
same time, B\"uttiker and his co-workers\cite{but4,but1,but2} have advanced 
a current conserving theory for the frequency dependent transport. This 
theory can be applied to discuss the non-linear behavior of mesoscopic 
samples. It has been recognized\cite{but3} that in non-linear coherent 
quantum transport, it is essential to consider the internal self-consistent 
potential in order to have the theory satisfy the gauge invariance
condition. This is a fundamental condition which requires that all 
physical properties predicted by a theory can not change if there is a 
global voltage shift. Recently, Christen and B\"uttiker\cite{but3} have 
investigated the rectification coefficient of a quantum point contact and 
the non-linear current-voltage characteristic of a resonant level in a 
double barrier structure using this theory of gauge invariant non-linear 
conductance. 

Clearly it is important and useful to further investigate nonlinear quantum
transport phenomena in coherent quantum conductors. In particular, detailed
predictions of nonlinear conductance of two-dimensional (2D) systems warrant
to be made because these systems can now be fabricated in many laboratories.
Unfortunately due to various technical difficulties, especially the
difficulty of evaluating a quantity called {\it sensitivity} 
(see below), so far the application of B\"uttiker's theory\cite{but3} has
largely been limited to quasi-1D systems. For 2D conductors, in general some
numerical calculation is needed and this has recently been carried out by
two of the authors\cite{wang1} to study the admittance of a T-shaped
conductor which is related to the {\it linear} order transmission function.
The nonlinear conductance for a 2D conductor, on the other hand, has been
investigated for a very special and exactly solvable model which is a
quasi-1D ballistic wire with a $\delta$-function impurity confined 
inside\cite{wang2,wang3}. Since it is exactly solvable\cite{wang2}, the
sensitivity can be computed in a closed form thereby overcoming the
technical difficulties associated with the theoretical formalism. 
To the best of our knowledge, this was the only explicit computation of the
weakly nonlinear conductance from the gauge invariant AC transport theory
for a 2D system where mode mixing is the most important characteristic.
However we note that in order to apply the theoretical formalism to a wide
range of 2D mesoscopic conductors, a more general numerical method must be
developed and various physical issues cleared. The purpose of this article
is to report our development of such a numerical method, and to investigate
the weakly nonlinear transport properties of a truly 2D conductor.

As we have noted from the previous investigation of the exactly solvable
model\cite{wang2}, for a geometrically symmetric system the second 
order non-linear conductance $G_{111}$ must be zero from a general
argument (see below). Hence the non-linear effect, {\it i.e.} a
non-zero $G_{\alpha \beta \gamma}$, obtained in Ref. \cite{wang2} 
is a delicate effect of the asymmetric scattering boundary\cite{but6}. 
Such an asymmetry is brought about when the $\delta$-function scatterer
is not located at the center of the scattering volume\cite{wang2}.
Already, very interesting and physically revealing behavior of the local
current response (the sensitivity) has been found. In this work, on the 
other hand, we shall focus on a much more general situation by
investigating the 2D conductor depicted in Fig. (1) where the scattering
volume is defined by the shaded area. The two leads with width $W$ 
are assumed to extend far away from the scattering volume. The shape 
of the side stub is controlled by the parameter $H$ as shown in Fig. (1), 
hence various different 2D systems can be generated by varying $H$. 
For $H=W$, the scattering volume is a geometrically asymmetric system, 
but the asymmetry is only due to the {\it asymmetric} locations of 
the scattering volume boundary.  For this case we thus expect that
the physics should be similar to those obtained in Ref. \cite{wang2}. 
For $H=2W$, the scattering volume becomes a geometrically symmetric 
system where $G_{\alpha \beta \gamma}$ must be zero. For other values 
of $H$ between $W$ and $2W$, the scattering volume is intrinsically 
asymmetric. By varying $H$, we shall study the crossover behavior of 
$G_{111}$ between the symmetric and asymmetric situations. 

Our results show that the external (the internal) response of the second 
order non-linear conductance changes sign from negative (positive) to 
positive (negative) near a quantum resonant point. The cancellation of 
the external and internal responses results in a much smaller second 
order non-linear conductance $G_{111}$, {\it i.e.}, $G_{111}$ is 
one order of magnitude smaller than the external or internal response. 
The behavior of $G_{111}$ is non-monotonic when changing 
the parameter $H$ in the range $W\leq H\leq 2W$: this is because $G_{111}$
is very small at $H=W$ as it is solely due to the asymmetric scattering 
boundary, it increases as $H$ is increased, and it is zero at $H=2W$. 
Another result of our analysis concerns the gauge invariant condition
$\sum_{\gamma} G_{\alpha \beta \gamma}=0$. It turns out that for systems with 
a finite scattering volume as those of any numerical calculations, if the 
global partial density of states (GPDOS) is computed from the 
{\it energy} derivatives of the scattering matrix, it was found\cite{wang2} 
that a correction term must be added to satisfy the gauge invariant condition. 
For the exactly solvable model studied in our previous work\cite{wang2}, 
this correction term has been derived\cite{wang2} analytically. 
We shall examine this effect for the conductor studied here.

The paper is organized as follows.  In the next section we shall
briefly review the gauge invariant theory for non-linear transport developed
by B\"uttiker\cite{but1}. The method of calculating various
quantities needed for non-linear conductance and our results are presented in
sections 3 and 4. The last section serves as a brief summary.

\section{The formalism}

The gauge invariant formalism of nonlinear transport has been developed and
clearly discussed in Ref. \cite{but3} and we refer interested reader to 
the original work. In this section, we shall outline the main steps of the
application of this formalism for our calculation. For a multi-probe 
mesoscopic conductor, the current through probe $\alpha$ is given 
by\cite{but1,but3}
\begin{equation}
I_{\alpha} = \frac{2e}{h} \sum_{\beta} \int dE f(E-E_F-eV_{\beta})
A_{\alpha \beta}(E, \{V_{\gamma}\}) \\,
\label{e1}
\end{equation}
where $f(E)$ is the Fermi distribution function, and  
\begin{equation}
A_{\alpha \beta}(E, \{V_{\gamma}\}) = Tr[{\bf{1}}_{\alpha} \delta_{\alpha 
\beta} - {\bf s}^{\dagger}_{\alpha \beta}(E, \{V_{\gamma}\}) {\bf s}_{\alpha
\beta}(E, \{V_{\gamma}\})]
\label{e2}
\end{equation}
are the screened (negative) transmission functions which are expressed in 
terms of the scattering matrix ${\bf s}_{\alpha \beta}$. For weakly
non-linear transport, Eq.(\ref{e1}) can be expanded with respect to the
voltages $V_{\beta}$.  For a conductor which has only two probes, up to 
the second order nonlinear term such an
expansion leads to the following equation\cite{but3,wang2},
\begin{equation}
I_1 = G_{11} (V_1 -V_2) + G_{111} (V_1 -V_2)^2\ \ \ ,
\label{iv}
\end{equation}
where $G_{11}$ is the usual linear conductance and $G_{111}$ is the second
order nonlinear conductance which we wish to compute for the conductor of
Fig. (1). 

It can further be proven\cite{but3} that $G_{\alpha\beta\gamma}$
is the sum of two terms which are the external and internal contributions:
\begin{equation}
G_{\alpha \beta \gamma} = G^e_{\alpha \beta \gamma}+ 
G^i_{\alpha \beta \gamma} \ \ 
\label{e4a}
\end{equation}
where the external contribution can be obtained using the free electron
scattering theory by evaluating the energy derivatives of the 
scattering matrix,
\begin{equation}
G^e_{\alpha \beta \gamma} = \frac{e^2}{h} \int dE (-\partial_E f)
e \partial_E A_{\alpha \beta} \delta_{\beta \gamma}\ \ .
\label{e5}
\end{equation}
The internal contribution, on the other hand, is much more difficult to
obtain because it depends on the potential derivatives of the
scattering matrix,
\begin{equation}
G^i_{\alpha \beta \gamma} = \frac{e^2}{h} \int dE (-\partial_E f)
( \partial_{V_{\gamma}} A_{\alpha \beta} +
\partial_{V_{\beta}} A_{\alpha \gamma})\ \ .
\label{e6}
\end{equation}
The reason why this is difficult to evaluate is because when the voltage of a
probe $V_{\gamma}$ changes, the entire potential landscape of the scattering
volume will change accordingly through the electron-electron interactions.
Hence the internal contribution to the nonlinear conductance
can be obtained only after an interacting electron problem has been 
solved\cite{but3}. This is a very difficult task and so far has not been
successfully implemented in a numerical scheme. However if we can use 
the Thomas-Fermi linear screening model, which is more appropriate for 
metallic conductors, the internal contribution can be computed through the
evaluation of quantities called sensitivity and characteristic
potential\cite{but3}. It can be shown\cite{but3} that the potential 
derivative of the transmission function is given as,
\begin{equation}
\partial_{V_{\gamma}} A_{\alpha \beta} = 4\pi \int d^3 {\bf r} 
\eta_{\alpha \beta}({\bf r}) u_{\gamma}({\bf r})
\label{e11}
\end{equation}
where 
\begin{equation}
\eta_{\alpha \beta}({\bf r}) = \frac{1}{4\pi} \frac{\delta A_{\alpha
\beta}}{\delta U({\bf r})} = 
-\frac{1}{4\pi} Tr({\bf s}^{\dagger}_{\alpha 
\beta} \frac{\delta {\bf s}_{\alpha \beta}}{\delta U({\bf r})}+{\bf s}_{\alpha 
\beta} \frac{\delta {\bf s}^{\dagger}_{\alpha \beta}}{\delta U({\bf r})}) 
\label{e12}
\end{equation}
is called {\it sensitivity}\cite{gas1} which measures the local electric
current response to an external perturbation. $u_{\gamma}({\bf r})$ is the
characteristic potential which measures the variation of the potential
landscape of the scattering volume due to the perturbation\cite{but1}. 
Within the Thomas-Fermi screening model, it is given by 
\begin{equation}
u_{\gamma}({\bf r}) = \frac{dn({\bf r},\gamma)}{dE}/
\frac{dn({\bf r})}{dE} \ \ .
\label{uu}
\end{equation}
Here the local partial density of states (LPDOS) $dn({\bf r},\gamma)/dE$ 
is called the injectivity and is given by\cite{but5} the scattering
wavefunctions,
\begin{equation}
\frac{dn({\bf r},\gamma)}{dE} = \sum_n \frac{|\Psi_{\gamma n}({\bf r})|^2}{h 
v_{\gamma n}} \ \ \ ,
\label{inject}
\end{equation}
where $v_{\gamma n}$ is the electron velocity for the propagating channel
labeled by $n$, and $dn({\bf r})/dE =\sum_{\alpha} dn({\bf r}, \alpha)/dE$ 
is the total local density of states.

From the weakly nonlinear conductance formalism summarized above, several
observations are in order.  First, from an application of this formalism 
point of view, a crucial step is the evaluation of the sensitivity 
$\eta_{\alpha\beta}$ which depends on the functional derivative of the
scattering matrix with respect to the local potential variation. The latter
is caused by the external perturbation, {\it i.e.} the change of the
electrochemical potential at a lead. We are aware of two ways of calculating
the sensitivity\cite{gas1}. The first is to evaluate 
$\delta {\bf s}_{\alpha \beta}/\delta U$ directly by introducing 
a $\delta$-function of infinitesimal strength $\delta U$ inside the 
scattering region. Alternatively, one
can calculate it using the retarded Green's function. For a 2D system, in
general the Green's function can not be obtained explicitly except in
very special cases such as that studied in Ref. \cite{wang3},
hence we shall use the first method by directly computing the sensitivity. 
As a second observation which is physically important\cite{but6}, we can
discuss the general behavior of the nonlinear conductance $G_{111}$.
From Eq. (\ref{iv}), for a symmetric scattering volume with scattering 
potential $U(x,y) = U(-x,y)$ where $x$ is the propagation direction, we 
must have $-I_1$ if $V_1$ and $V_2$ are interchanged. Hence we conclude 
that for a symmetric scattering volume there is no quadratic terms, 
{\it i.e.}, $G_{111}=0$. On the other hand, in general $G_{111}\neq 0$ 
for geometrically asymmetrical systems.  Finally, due to the current 
conservation and the gauge invariance condition, namely the entire physics 
is independent of a global voltage shift, it is not difficult to 
prove\cite{but1,but3,but5}
\begin{equation}
\sum_{\alpha} G_{\alpha \beta \gamma} = \sum_{\beta} G_{\alpha
\beta \gamma} = \sum_{\gamma} G_{\alpha \beta \gamma} = 0\ \ \ .
\label{e7}
\end{equation}
Our results will allow a direct confirmation of this equation.

\section{Numerical method}

There are several ways to solve the scattering matrix of 2D ballistic
conductors, such as the mode-matching method\cite{schult}, the recursive
Green's function method\cite{lee,baranger,datta}, 
and the finite-element method\cite{lent,ywang}.
However we found that all these methods are not particularly easy to apply 
here because we need not only the scattering matrix, but also the
sensitivity $\eta_{\alpha\beta}$.  For this purpose, we found that a 
technique for computing scattering matrix which is developed in 
Ref. \cite{sheng} is quite useful and we shall discuss, in some detail, 
our numerical procedure for finding $\eta_{\alpha\beta}$.

In particular, we construct a global scattering matrix using 
the mode-matching method of Ref. \cite{sheng}. For a scattering volume 
which is not uniform along its longitudinal direction, 
we divide it into a number of uniform sections, {\it e.g.}, 
the asymmetric cavity as shown in Fig. (1) can be divided 
into four uniform sections. The scattering matrix associated 
with the $n$-th section ${\bf S}_n$ is the composition of 
two individual scattering matrices ${\bf S}_n^f$ and ${\bf S}_n^i$, 
{\it i.e.}, ${\bf S}_n={\bf S}_n^f\otimes{\bf S}_n^i$ 
where $\otimes$ is the operator which denotes the 
composition of two scattering matrices\cite{tamura}.
Here ${\bf S}_n^f$ describes the free propagation from the 
left end of the $n$-th section to its right end and is given by
\begin{equation}
{\bf S}_n^f(L_n)=\left [\begin{array}{cc}
{\bf 0}&{\bf P_n}\\
{\bf P_n}&{\bf 0}\\
\end{array}\right ]
\end{equation}
where ${\bf P_n}$ is a diagonal matrix with elements $({\bf P}_n)_{mm}=
e^{ik_n^mL_n}$, $k_n^m$ is the longitudinal wave number for the $m$-th
mode and $L_n$ is the length of the $n$-th section. The scattering 
process at the interface between two adjacent sections 
(the $n$-th and $(n+1)$-th sections) is described by ${\bf S}_n^i$. 
Care must be taken when matching the wavefunctions of two sections with 
different widths at the section boundary. If the width of the $n$-th 
section $W_n$ is not greater than $W_{n+1}$, we have 
\begin{equation}
{\bf S}_n^i=
\left [\begin{array}{cc}
-{\bf C}^T&{\bf I}\\
{\bf K}_n&{\bf C}{\bf K}_{n+1}\\
\end{array}\right ]^{-1}
\left [\begin{array}{cc}
{\bf C}^T&-{\bf I}\\
{\bf K}_n&{\bf C}{\bf K}_{n+1}\\
\end{array}\right ]
\end{equation}
where ${\bf K}_n$ is a diagonal matrix with diagonal element $k_n^m$, and
${\bf I}$ is a unit matrix. ${\bf C}$ is a matrix which denotes the coupling
between the transverse modes in the two sections and its elements are given
by $C_{ij}=\langle\phi_n^i|\phi_{n+1}^j\rangle$ where $\phi_n^i$ is the ith
transverse mode in the nth section. ${\bf C}^T$ is the transpose of the
matrix ${\bf C}$. On the other hand if $W_n>W_{n+1}$, we have
\begin{equation}
{\bf S}_n^i=
\left [\begin{array}{cc}
-{\bf I}&{\bf C}\\
{\bf C}^T{\bf K}_n&{\bf K}_{n+1}\\
\end{array}\right ]^{-1}
\left [\begin{array}{cc}
{\bf I}&-{\bf C}\\
{\bf C}^T{\bf K}_n&{\bf K}_{n+1}\\
\end{array}\right ]
\end{equation}
Once the scattering matrix for each section is known, the global scattering
matrix can be easily constructed by the composition of all the individual
scattering matrices,
\begin{equation}
{\bf S}={\bf S}_1\otimes{\bf S}_2\otimes\ldots\otimes{\bf S}_{M-1}
\label{total}
\end{equation}
where $M$ is the total number of sections.

It should be noted that the global scattering matrix ${\bf S}$ calculated 
this way is different from the standard one\cite{baranger} which only
connects the outgoing wave to the incoming wave. Here, the global
scattering matrix ${\bf S}$ involves the non-propagating channels as well
and does not satisfy the unitarity condition. In order to obtain a physical 
scattering matrix which we denote by the lower case ${\bf s}$, we first
rewrite global ${\bf S}$ in the form of $2\times2$ subblocks and obtain four
sub-matrices ${\bf S}_{ij}$ where $(i,j=1,2)$. Then for each sub-matrix
${\bf S}_{ij}$ we build a new matrix ${\bf s}_{ij}$ constructed by the first
$N_{0}$ rows and columns of ${\bf S}_{ij}$, where $N_0$ is the number of
propagating channels. By writing the four newly constructed matrices 
in the form of $2\times2$ blocks, a $2N_0$-dimensional scattering 
matrix ${\bf s}$ is obtained which is the true scattering matrix that
connects the outgoing wave to the incoming wave. In order to obtain a
unitary scattering matrix ${\bf s}$, one should further take a unitary
transformation.
\begin{eqnarray}
&{\bf s}={\bf A}{\bf s}{\bf A}^{-1}&
\nonumber\\
&{\bf A}=
\left [
\begin{array}{cc}
{\bf V}&{\bf 0}\\
{\bf 0}&{\bf V}
\end{array}
\right ]&
\end{eqnarray}
where ${\bf V}$ is a $N_0$-dimensional diagonal matrix with diagonal
element $\sqrt{k_n}$.

The above procedures can easily be modified to compute the sensitivity
$\eta_{\alpha\beta}({\bf r})$. For this purpose, we shall make use of a
$\delta$-function impurity to calculate the functional derivatives of the
scattering matrices $\delta {\bf s}_{\alpha\beta}/\delta U({\bf r})$. 
This is achieved as follows. We put a $\delta$-function impurity 
with infinitesimal strength $\gamma$,
$V({\bf r})=\gamma\delta({\bf r}-{\bf r}_0)$, at arbitrary positions
${\bf r}={\bf r}_0$ in the scattering volume. We then calculate the scattering
matrix ${\bf s}_{\alpha\beta}$ as a function of $\gamma$. Finally 
we use a five-point numerical derivative to evaluate 
$\delta {\bf s}_{\alpha \beta}/
\delta U({\bf r}) \equiv \partial {\bf s}_{\alpha\beta}/\partial
\gamma |_{\gamma=0}$.  With this result we can obtain the sensitivity 
from Eq. (\ref{e12}).

The scattering matrix can still be obtained using the approach discussed 
above even including the $\delta$-function impurity. In fact it has been 
derived in the presence of a magnetic field by Tamura and Ando\cite{tamura}. 
Here we give the expression in the absence of the field\cite{ferry}. 
Suppose the $\delta$-function impurity is located in the $n$-th section at 
position ${\bf r}_0=(x_0,y_0)$, where $x_0$ and $y_0$ is the distance from 
the left and bottom boundary of the section respectively. The scattering 
matrix associated with this section is then given by
\begin{equation}
{\bf S}_n={\bf S}_n^f(x_0)\otimes{\bf S}_n^\delta\otimes{\bf S}_n^f(L_n-x_0)
\otimes{\bf S}_n^i\ \ .
\end{equation}
Here ${\bf S}_n^\delta$ describes the scattering process associated with the
$\delta$-function impurity and is given by\cite{ferry}
\begin{equation}
{\bf S}_n^\delta=
\left [\begin{array}{cc}
-{\bf I}&{\bf I}\\
i{\bf K_n}-{\bf \Gamma}&i{\bf K_n}\\
\end{array}\right ]^{-1}
\left [\begin{array}{cc}
{\bf I}&-{\bf I}\\
i{\bf K_n}+{\bf \Gamma}&i{\bf K_n}\\
\end{array}\right ]
\end{equation}
where the matrix ${\bf \Gamma}$ describes the mode-mixing effect due to the
$\delta$-function impurity and its matrix elements are given by $\Gamma_{pq}=
2\gamma\sin(p\pi y_0/W_n)\sin(q\pi y_0/W_n)/W_n$ with $W_n$ being the 
width of the section.  With the $\delta$-function included this way, we can
again apply our method, Eq. (\ref{total}), to compute the scattering matrix
${\bf s}={\bf s}(\gamma)$ and complete the numerical derivatives discussed
in the last paragraph.

To end this section, we briefly mention two other points.  First, the
characteristic potential as given by Eq. (\ref{uu}) is evaluated using the
scattering wavefunction which can be calculated in two ways, directly or 
indirectly. One can directly compute the wavefunction using the mode 
matching method\cite{schult} or finite element method\cite{lent,ywang}. Or 
one can compute the wavefunction ($|\Psi|^2$) indirectly by computing the 
local partial DOS called emissivity defined as\cite{but5}

\begin{equation}
\frac{dn(\alpha,{\bf r})}{dE} = -\frac{1}{4\pi i} \sum_{\beta} 
Tr({\bf s}^{\dagger}_{\alpha \beta} 
\frac{\delta {\bf s}_{\alpha \beta}}{\delta eU({\bf r})}-{\bf s}_{\alpha \beta} 
\frac{\delta {\bf s}^{\dagger}_{\alpha \beta}}{\delta eU({\bf r})}) 
\label{emiss}
\end{equation}
The microreversibility of the scattering matrix implies, 

\begin{equation}
\frac{dn(\alpha,{\bf r},B)}{dE} = \frac{dn({\bf r}, \alpha, -B)}{dE}
\label{micro}
\end{equation}
where $B$ is the magnetic field. We will use Eqs.(\ref{inject}),(\ref{emiss}), 
and (\ref{micro}) to compute the wavefunction $|\Psi|^2$ since we have
to compute the scattering matrix or $\delta {\bf s}/\delta U({\bf r})$ 
anyway for the sensitivity. Second, the energy derivatives of the
transmission function, which determines the external response contribution to
the second order nonlinear conductance, is evaluated using a five point 
numerical difference.  These procedures are the same as that of our earlier
work\cite{wang1}.

\section{Results}

As a first result we plot in Fig. (2) the sensitivity $\eta_{11}({\bf r})$ 
at three different positions inside the scattering volume,
${\bf r}_1 = (0.5W,0.5W)$, ${\bf r}_2 = (0.5W, 1.25W)$, and
${\bf r}_3 = (1.25W, 0.625W)$, as a function of the normalized electron 
momentum $k_FW/\pi$ where $k_F$ is the electron Fermi wave number. 
These results are for an asymmetric system (see Fig. (1)) with the
parameter $H=1.25W$. As discussed in Section 2
$\eta_{\alpha \beta}$ appears naturally in the theoretical formalism, and 
it essentially describes the local (internal) electric current response of 
the scattering problem when there is a small local potential change. It
is related to the real part of the diagonal elements of the scattering
Green's function\cite{gas1}. From Fig. (2), we see that different positions
inside the scattering volume have quite different internal responses in
terms of the peak heights of the apparent resonance behavior.  On the other
hand, the peak positions occur at the same electron energies given by 
$k_FW/\pi$ for $\eta_{11}$ at all three positions.  We have checked (see
below and Fig. (5b)) that the peak positions also coincide with those of
the conductance. Hence we may conclude that the local current response 
can have sharp changes, from positive values to negative values, 
across the energy of a resonance which also mediates a resonance
transmission. For the two positions ${\bf r}_1$ and ${\bf r}_2$ 
which are located in the left part of the cavity, the shapes of the
sensitivities are more similar to each other.  This is to be compared
with that of the position ${\bf r}_3$ which is located 
in the right part of the cavity.  The differences are evident from the three
curves of Fig. (2).

To get a more intuitive picture of the spatial dependence of the 
sensitivity, in Fig. (3) we plot $\eta_{11}({\bf r})$ in the entire 
scattering volume for two different values of the electron momentum 
($H=1.25W$).  First is for $k_FW/\pi=1.715$ which is off resonance,
while the second is for $k_FW/\pi=1.795$ which is on resonance.
From the lower panel of Fig. (3) which corresponds to the resonant 
energy, the behavior of $\eta_{11}$ is reminiscent of a standing wave which
is in accordance to our usual picture of a quantum resonance. The positions
${\bf r}_1$ and ${\bf r}_2$ of Fig. (2) are located at 
a peak of the sensitivity profile while ${\bf r}_3$ is at a valley. This
explains why in Fig. (2) we observe the large resonant peak at 
${\bf r}_1$ and ${\bf r}_2$ but not at ${\bf r}_3$. When off resonance,
the upper panel of Fig. (3) shows less regular patterns for $\eta_{11}$.
Hence the local current response can behave regularly or less regularly 
depending on the electron Fermi energy being on or off a quantum resonance
of the scattering cavity. In comparison, for both 1D and 2D scattering problems 
involving a $\delta$ potential barrier, the sensitivity has been derived 
analytically in Ref. \cite{gas1} and \cite{wang2}. There, $\eta_{11}$ 
shows strong spatial regular oscillations.

When the geometry parameter $H=W$, the conductor becomes a T-shaped
junction. The upper panel of Fig. (4) 
shows $\eta_{11}$ for this situation at $k_FW/\pi=1.325$.
The lower panel of Fig. (4) plots the characteristic potential
$2u_1({\bf r})-1$ for this case.  The quantity $2u_1({\bf r})-1$ is
interesting because it can easily be shown\cite{but3}, using Eqs.
(\ref{e4a}), (\ref{e5}), (\ref{e6}) and applying the gauge invariance
condition (\ref{e7}), that the nonlinear conductance can be re-written as  
\begin{equation}
G_{\alpha \beta \gamma} = 4\pi\frac{e}{h} \int dE (-\partial_E f)
\int d^3 {\bf r} (\eta_{\alpha\beta} u_{\gamma}({\bf r}) + \eta_{\alpha\gamma}
u_{\beta}({\bf r}) - \eta_{\alpha\beta} \delta_{\gamma \beta}) \ \ .
\label{correct}
\end{equation}
Hence the quantity $2u_1({\bf r})-1$ appears naturally in this form of
$G_{111}$. From Fig. (4) it is clear that $\eta_{11}$ is symmetric 
and $2u_1-1$ is anti-symmetric along $x$ axis. As a result, $G_{111}$ will 
be zero for this T-shaped junction if the scattering volume is symmetric
due to the spatial integration of Eq. (\ref{correct}).  To systematically 
investigate the behavior of $G_{111}$ as the conductor shape changes from 
symmetric to asymmetric, we have calculated this quantity for several 
values of the geometric parameter $H$ at zero temperature: $H=W, 1.25W, 
1.5W$, and $1.75W$ (see Fig. (1)). Figs. (5a-5d) plot 
the DC-conductance $G_{11}$, the external and internal responses of the 
second order non-linear conductance, $G^e_{111}$\cite{foot1},
$G^i_{111}$, and $G_{111}$ as a function of the normalized electron 
momentum $k_FW/\pi$ for these configurations. 
Several interesting features have been observed.
First of all, the external (the internal) response of the 
second order non-linear conductance changes sign from negative (positive) to 
positive (negative) near the resonant point. This behavior is similar to
that of one-dimensional asymmetric double barrier resonant 
tunneling\cite{but3}. In that case 
$G_{111} = (e^3/h)(dT/dE)(\Gamma_2- \Gamma_1)/\Gamma$, where $T$ is the 
transmission coefficient, $\Gamma_i$ is the decay width of each barrier,
and $\Gamma=\Gamma_1+\Gamma_2$. Because of the term $dT/dE$, $G_{111}$
changes sign across the resonant point and hence can be negative. 
The cancellation of the external and internal responses results
in a much smaller $G_{111}$: one order of magnitude smaller than the
internal or external contribution alone. Secondly, for $H=W$ the asymmetry
of the scattering volume only comes from the location of the 
scattering volume boundary which is at our disposal, $G_{111}$ has the 
smallest values for all $H<2W$ studied. For $W<H<2W$, the conductor 
is intrinsically asymmetric and $G_{111}$ are larger. Also the resonance 
behavior of $G_{111}$ becomes substantially sharper as $H$ is increased.
While $G_{111}$ increases as $H$ increases from $H=W$, it eventually starts
to decrease after $H>1.5W$.  This is clearly seen from Fig. (5d) for which
$H=1.75W$.  This is because for larger $H$ the system approaches to the
symmetric conductor at $H=2W$, for which $G_{111}=0$ as discussed above.

So far, as noted in Ref. \cite{foot1}, we have computed the external
contribution $G_{111}^e$ of Eq. (\ref{e5}) using a procedure which employs
the gauge invariance condition. However if we directly use the right hand
side of Eq. (\ref{e5}) to compute $G_{111}^e$, the result will not be
accurate because the scattering volume is finite. In particular, as 
studied previously\cite{wang2} for a symmetric system $G_{111}$ will
be non-zero if calculated using the right hand side of Eq.(\ref{e5}). 
In terms of the gauge invariant condition, this will lead to a violation,
{\it i.e.} $G_{111}+G_{112}\neq 0$. While such an error is not relevant
if the scattering volume is very large, for our numerical calculations it
will generate incorrect conclusions since the scattering volume is always
finite. Hence, in order to use the right hand side to compute $G_{111}^e$,
a correction term is needed to preserve the gauge invariance. 

For a quasi-1D wire with a $\delta$ function impurity, this correction 
term has been derived\cite{wang2} analytically. For that situation, the
correction, $C$, consists of two terms\cite{wang2}:
\begin{equation}
C = \frac{|s_{12}|^2}{k^2_1} Re(s_{11}) +
Re(\sum_{n=2} \frac{b_1 |b_n|^2} {k_1 k_n} e^{ik_n (x_2-x_1)})\ \ .
\label{cor}
\end{equation}
where $k_n$ is the longitudinal momentum for the $n$-th mode with 
$k_n^2 = k_F^2 - (n\pi/W)^2$ and $E = \hbar^2 k_F^2/2m$, $x_1$ and $x_2$ 
are the coordinates for the scattering boundaries, and $b_n$ is the reflection 
scattering amplitude. Clearly, the first term is oscillating as the linear 
size of the scattering volume increases (due to $s_{11}$) and it is
only relevant near the edge of the first propagation threshold; the
second term is exponentially decaying to zero as the size of the volume 
increases and it comes solely from mode mixing and contributed by the 
evanescent modes.  Although this form of the correction term was derived 
from another system, it is nevertheless interesting to compare this 
formula for the conductor studied here.

In Fig. (6) we plot the correction term Eq. (\ref{cor}) together with
$G_{111}+G_{112}$ which were computed using the right hand side of Eq.
(\ref{e5}), for the case $H=W$. Eq. (\ref{cor}) was evaluated using the
scattering matrix elements obtained from our numerical calculation, and the
system size $(x_2-x_1)$ was specified as the length of our scattering
volume $2L+2W$ (see Fig. (1)). For this conductor with $H=W$, we expect a
good comparison with Eq. (\ref{cor}) which was derived for a
$\delta$-function scatterer inside a wire, because for both systems the
geometric asymmetry is solely due to the positions of the scattering volume
boundary.  Other than that, these systems are actually {\it symmetric} 
with respect to the scattering potential.  Indeed, Fig. (6) clearly shows
that there is essentially no difference between Eq. (\ref{cor}) and our
numerical data when the scattering volume is large ($L=4W$).
On the other hand, for a conductor with $H=1.5W$ which has an intrinsically
asymmetric scattering volume, the comparison is qualitative as shown
in Fig. (7). However the trend of the two curves are still similar.
We may thus conclude that for the gauge invariant condition,
the correction term to the external contribution Eq. (\ref{e5}) has a form
with the same nature as that of Eq. (\ref{cor}) above.  

\section{Summary}

In this work we have developed a numerical technique based on a scattering
matrix to compute the weakly nonlinear conductance. This technique is
particularly useful for conductors whose scattering volume can be naturally
divided into several regions. The most difficult step is the evaluation of
the local electric current response, namely the sensitivity.  We have
reported how to obtain this quantity numerically, thus further
investigations of the interesting nonlinear conductance problem can be
carried out using our numerical method.  We have found that sensitivity
behaves differently when the transport energy is on or off resonance.  The
former leads to standing-wave type spatial dependence, while the latter is
behaving in a less regular fashion.  In all cases, the sensitivity shows
spatial oscillating pattern, which is similar to those known from exact
calculations for 1D models.

The nonlinear conductance can be non-zero only for geometrically 
asymmetric systems.  The asymmetry can be introduced in two ways.  
The first is through the intrinsically asymmetrical shape of a conductor 
such as those of Fig. (1) with $W<H<2W$. The other, which is a trivial
asymmetry, is through the asymmetrical location of the scattering volume
boundary, {\it e.g.} the case $H=W$. We discover that the intrinsical 
asymmetry leads to much larger nonlinear conductance than the other case.  
Furthermore, for the symmetrical scattering junction but with asymmetrical
location of the boundary ($H=W$ case), for large size $L$ the behavior of 
the gauge invariance condition agrees almost perfectly with Eq. (\ref{cor}) 
which was derived from a completely different system but also with the
asymmetry introduced by the location of the scattering volume boundary
only.  On the other hand, such an agreement is less perfect for the
intrinsically asymmetrical system.  Hence we may conclude that the nonlinear
conductance behaves in quite different manner depending on how the asymmetry
is introduced.

The sign of the nonlinear conductance can be positive or negative.  Very
sharp variations of this quantity is discovered at quantum resonances for
the conductor studied here, where such resonances are marked by sharp
reductions of the linear conductance $G_{11}$. Hence near a resonance, the
electric current may actually decrease for an increasing voltage difference
by Eq. ({\ref{iv}) since $G_{111}$ is negative.  Such a behavior is precisely
the expected nonlinear conduction characteristic, and up to the second order in
voltage difference our results can provide a prediction.  Clearly, as the
voltage difference becomes large, even higher order conductances must be
included in order to have a meaningful prediction of the nonlinear 
I-V curve.

\section*{Acknowledgments}

We gratefully acknowledge support by a research grant from the Croucher 
Foundation, a RGC grant from the Government of Hong Kong under grant number 
HKU 261/95P, the NSERC of Canada and FCAR of Qu\'ebec.  We thank the Computer 
Center of the University of Hong Kong for computational facilities.

\newpage

$^*$ permanent address: 
{\it Institute of Semiconducters,\\
Academia Sinica, \\
P. O. Box 912,\\
Beijing, P. R. China}

\newpage
\section*{Figure Captions}

\begin{itemize}
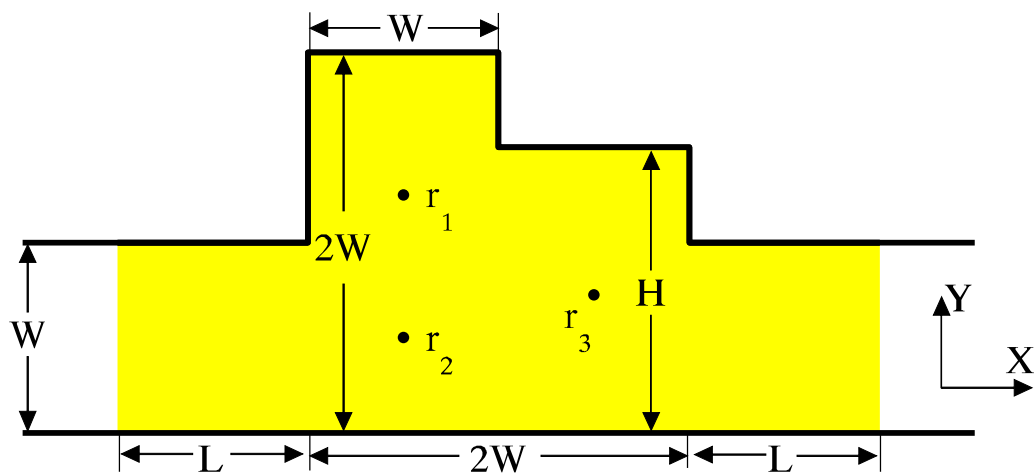
\item[{Fig. 1}] Schematic view of an asymmetric cavity (the shaded area) 
embedded in a quantum wire.

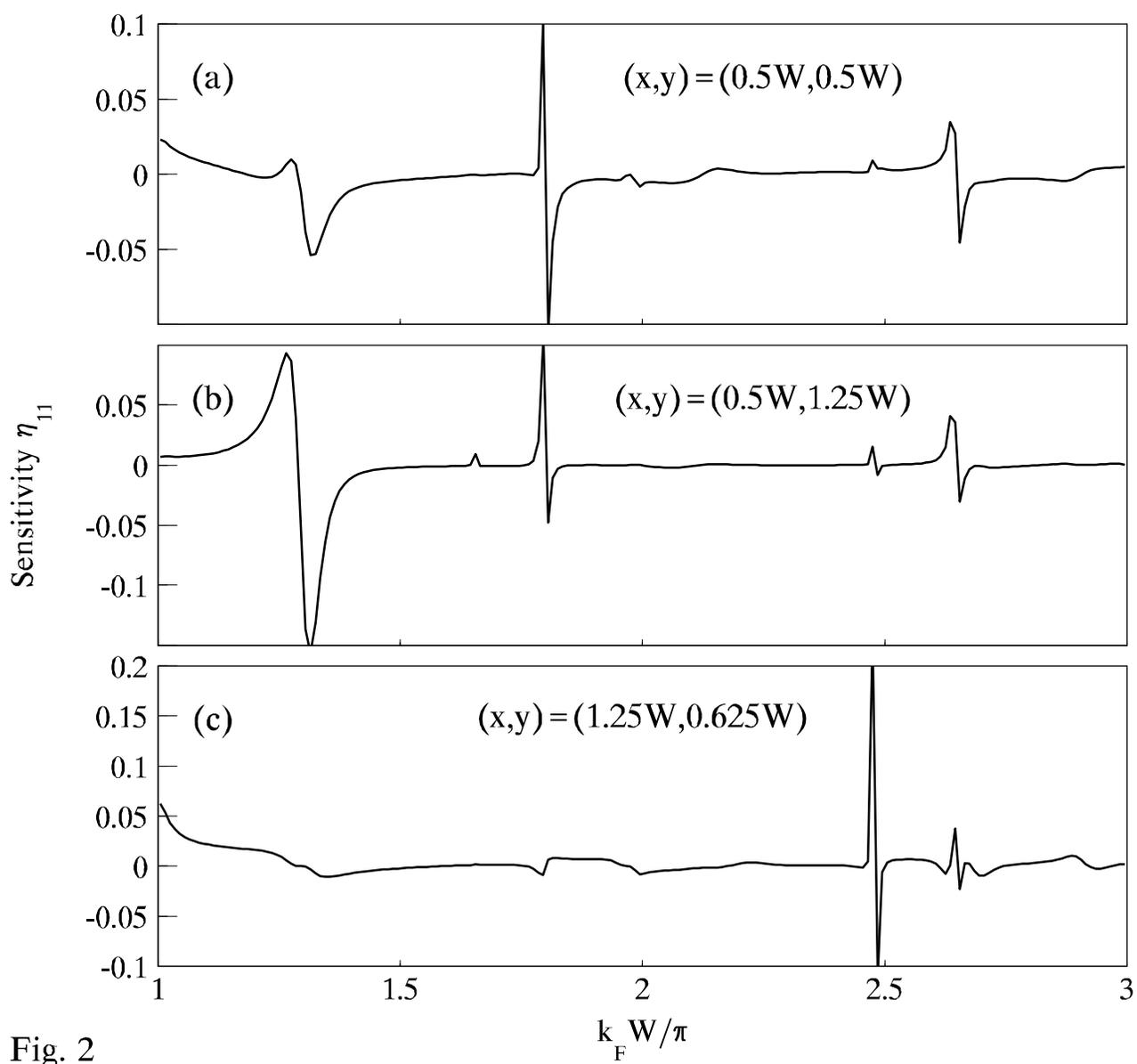
\item[{Fig. 2}] The sensitivity $\eta_{11}$ at three different positions,
(x, y)=\{(0.5W, 0.5W), (0.5W, 1.25W), (1.25, 0.625W)\},
as a function of the normalized electron momentum $k_FW/\pi$ for $H=1.25W$. 

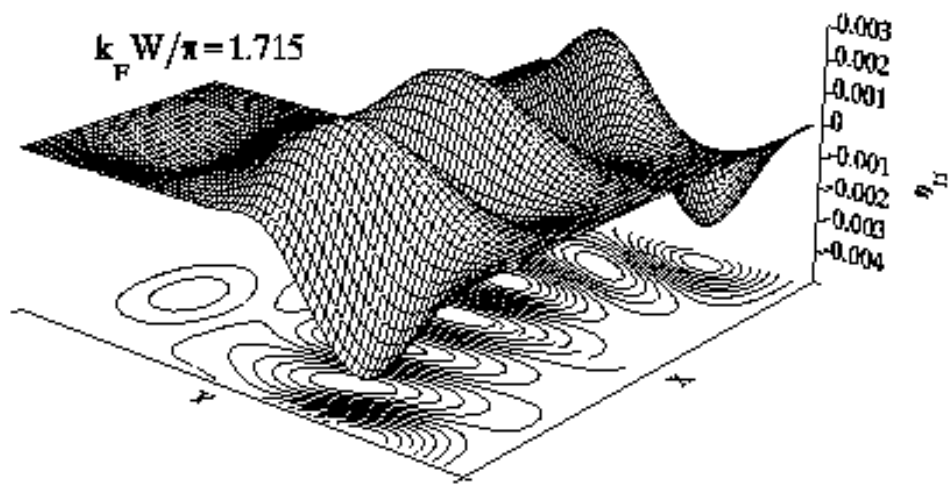
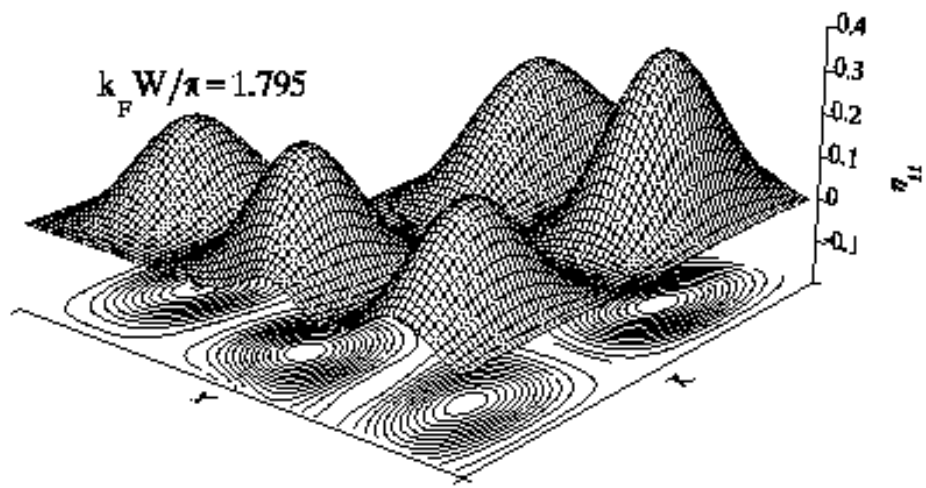
\item[{Fig. 3}] Three dimensional view of the sensitivity $\eta_{11}$ for 
$H=1.25W$ at two different values of the electron momentum $k_FW/\pi=1.715$ 
and $k_FW/\pi=1.795$.

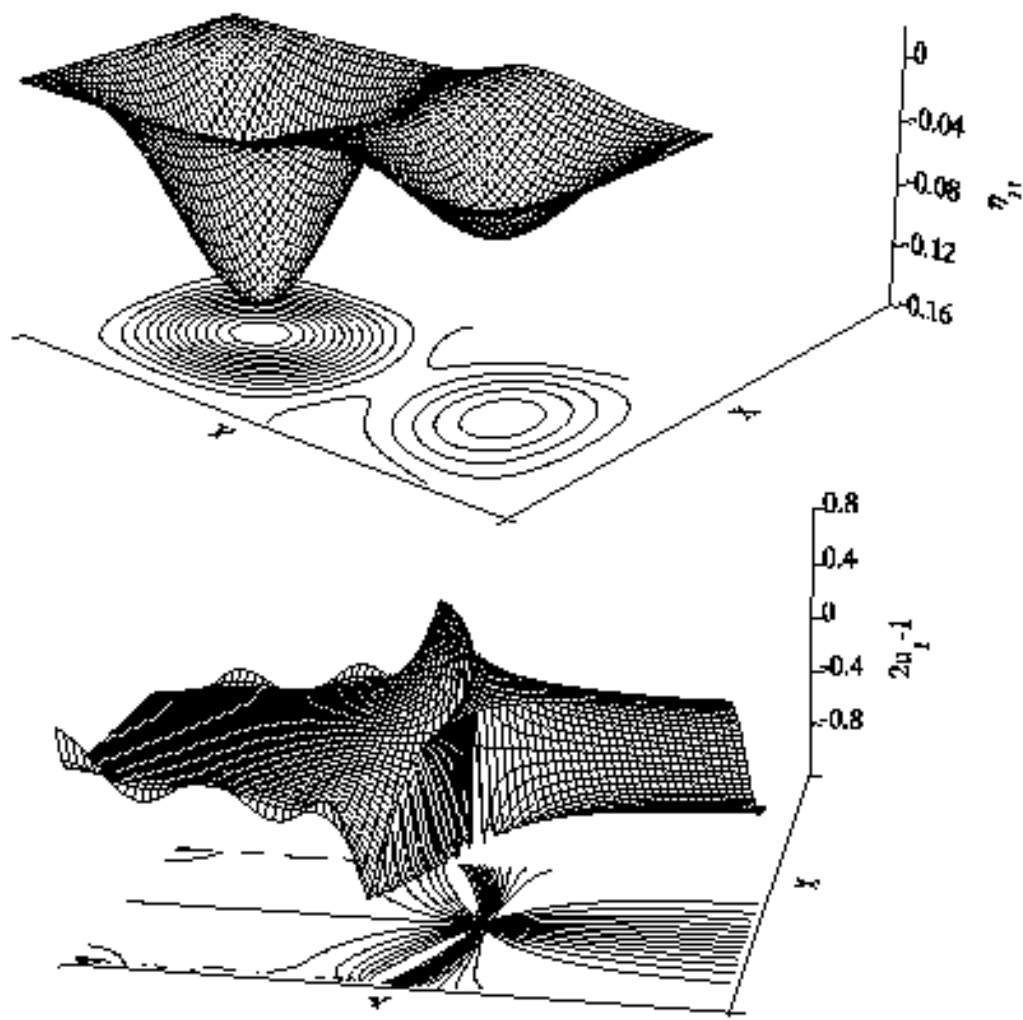
\item[{Fig. 4}] Three dimensional view of the sensitivity $\eta_{11}$ and
the characteristic potential $2u_1-1$ for a symmetric T-shaped cavity
at $k_FW/\pi=1.325$ and $H=W$. 

\item[{Fig. 5}] The DC-conductances $G_{11}$ and the leading order nonlinear
terms $G_{111} $ as a function of the normalized electron momentum
$k_FW/\pi$, solid lines for $G_{111}^e$ and dotted lines for
$G_{111}^i$. (a) $H=W$, (b) $H=1.25W$, (c) $H=1.5W$, (d) $H=1.75W$.
Here $E_1$ is the threshold of the first subband defined as $E_1 = \hbar^2
\pi^2/(2m W^2)$. 

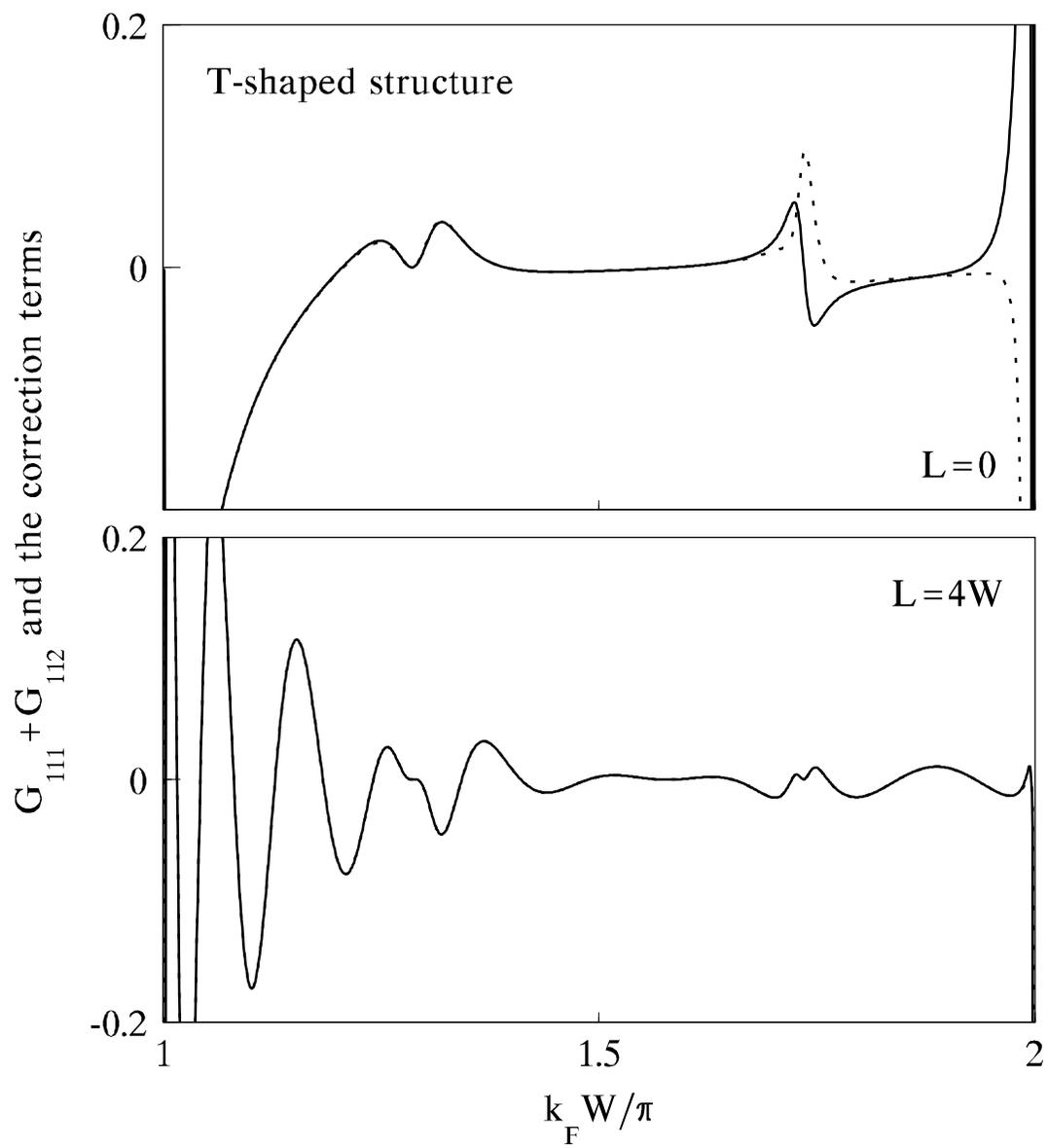
\item[{Fig. 6}] $G_{111}+G_{112}$ (solid line) and the correction term 
according to Eq.(\ref{cor}) (dotted line) versus momentum for the
T-shaped cavity ($H=W$). Upper panel: $L=0$. Lower panel: $L=4W$.

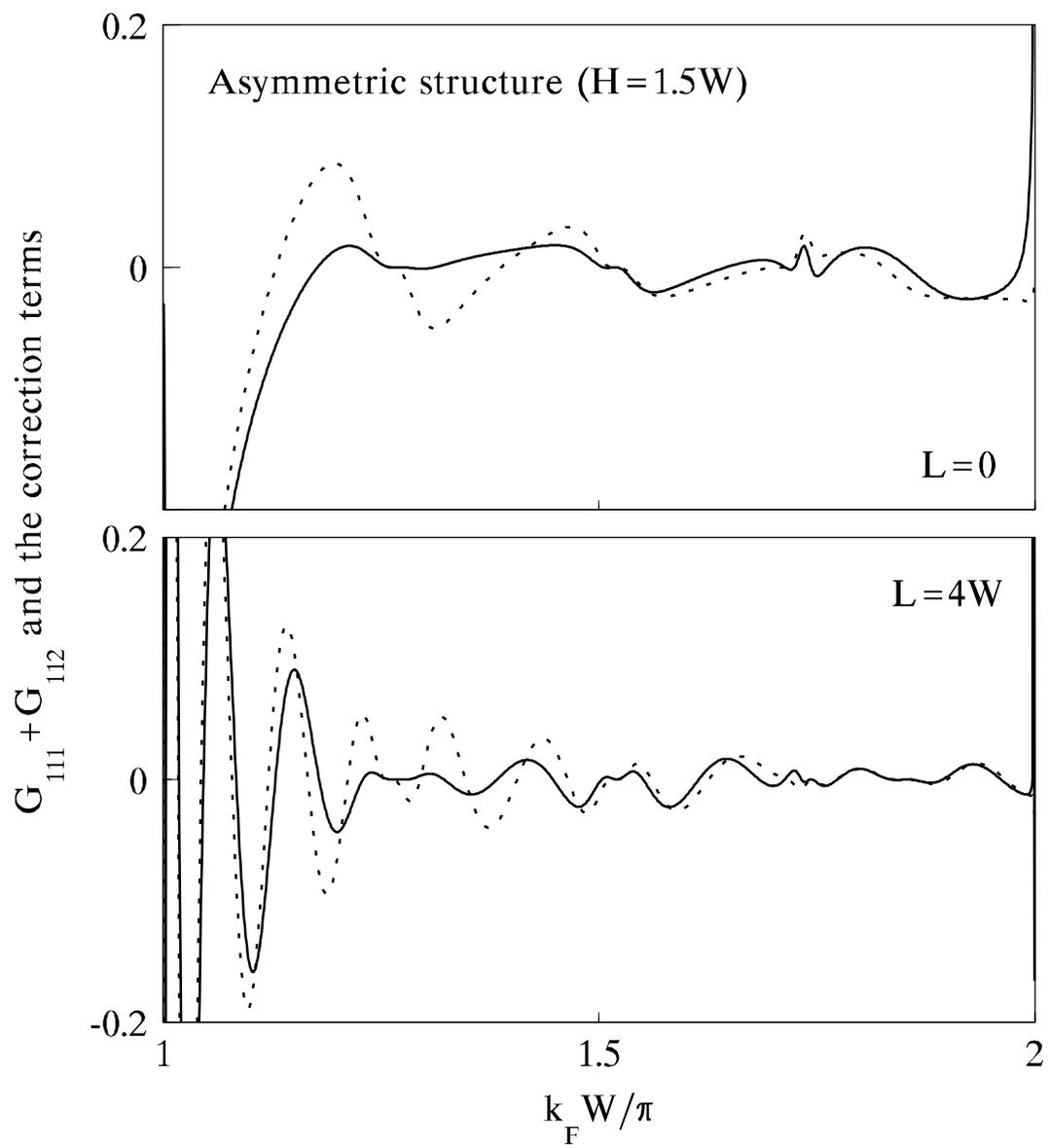
\item[{Fig. 7}] $G_{111}+G_{112}$ (solid line) and the correction term 
according to Eq.(\ref{cor}) (dotted line) versus momentum for an
asymmetric structure ($H=1.5W$). Upper panel: $L=0$. Lower panel: $L=4W$.

\end{itemize}

\begin{thebibliography}{00}
\bibitem{foot1}
The $G_{111}^e$ of Fig. (5) was computed by replacing the energy derivatives
of Eq. (\ref{e5}) with a potential derivative using the gauge invariance
condition. This procedure can be proven using the original formula in 
Ref. \cite{but5,leavens} and numerically verified in Ref. \cite{wang2}.
\end{thebibliography}
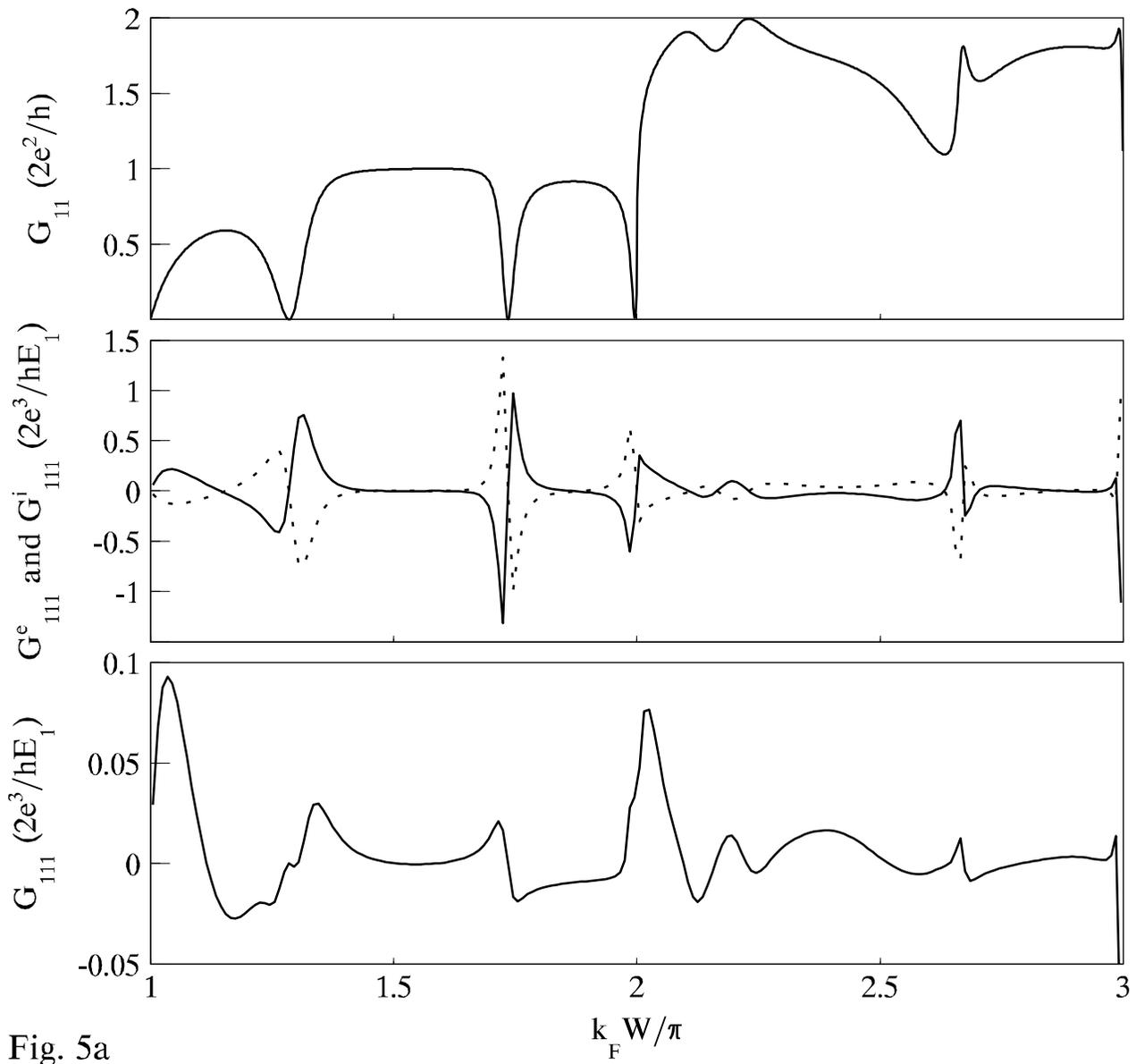
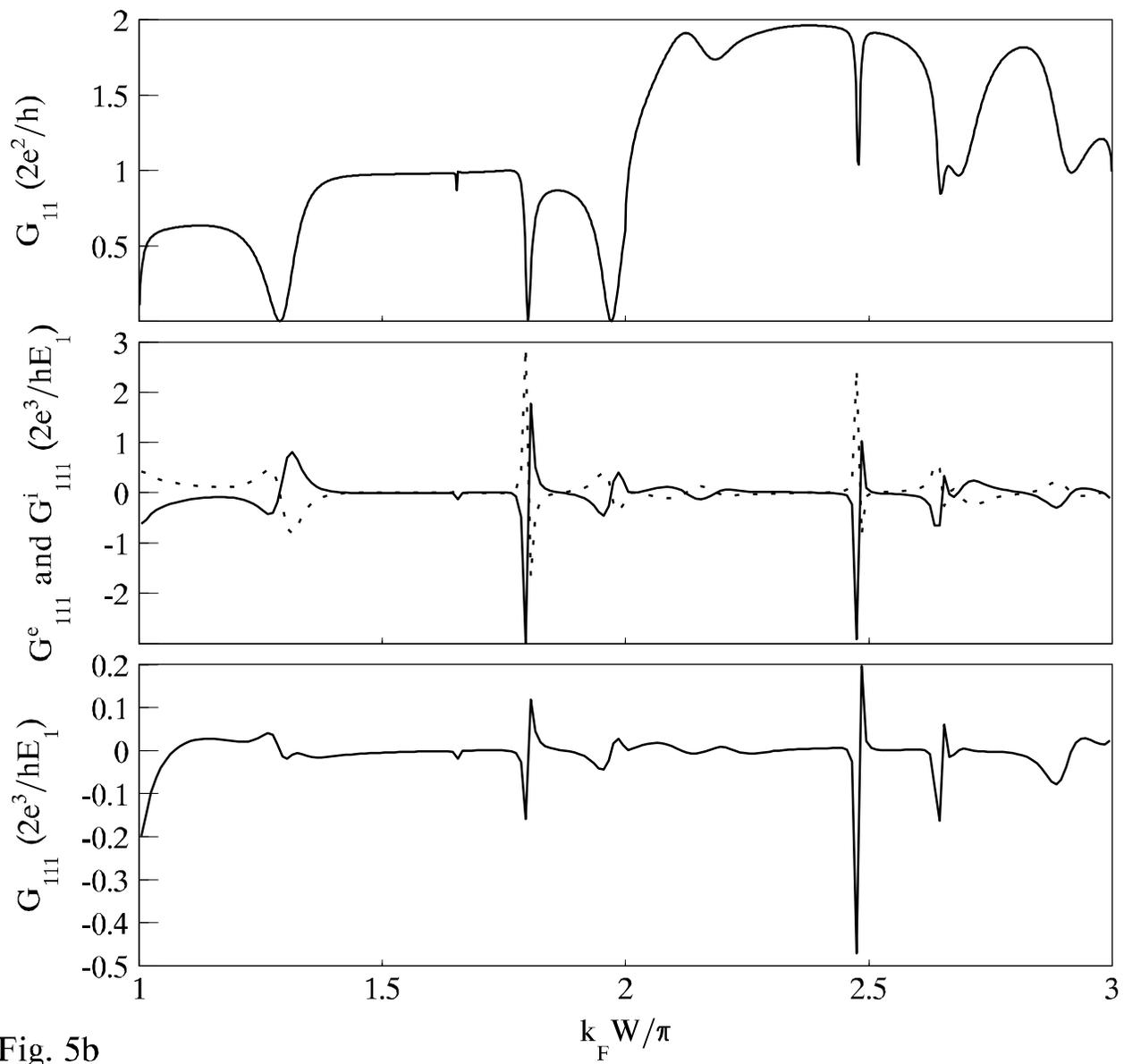
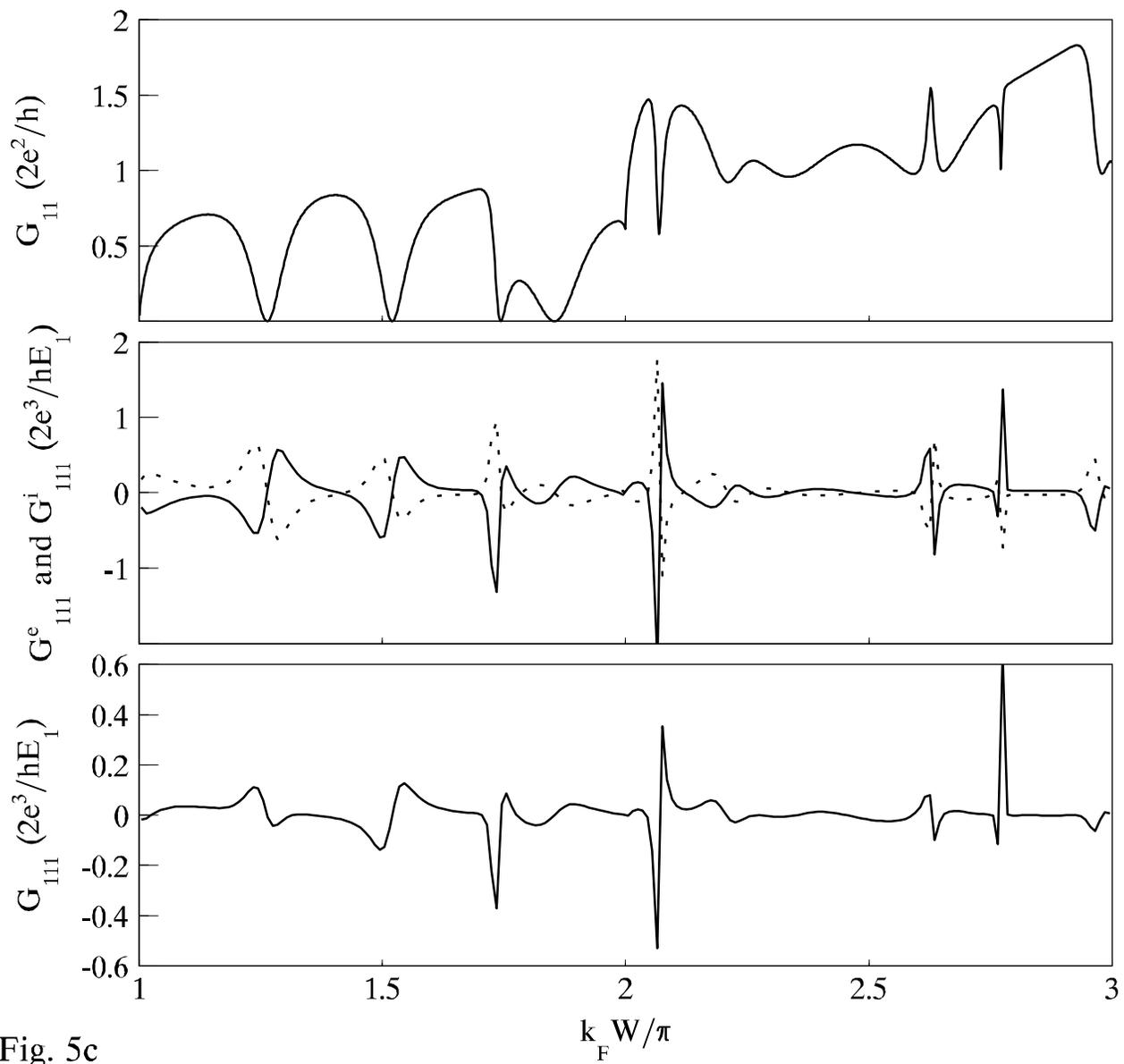
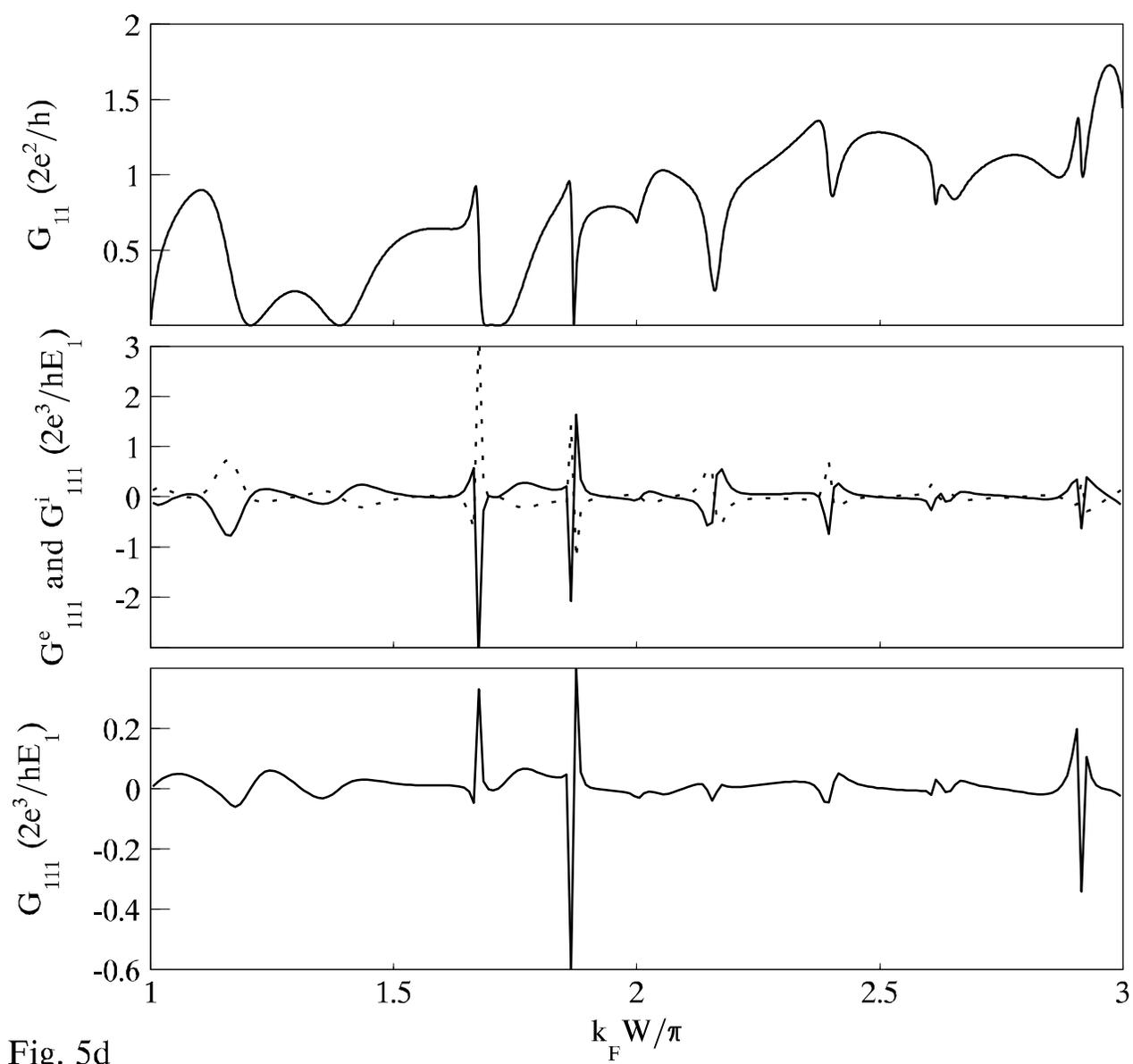
\end{document}